\documentstyle[11pt,moriond,epsfig]{article}
\input psfig.tex

\def\be{\begin{equation}}
\def\ee{\end{equation}}
\def\bea{\begin{eqnarray}}
\def\eea{\end{eqnarray}}

\begin{document}
\vspace*{4cm}
\title{COSMIC STRINGS IN A UNIVERSE WITH NON-CRITICAL MATTER DENSITY}

\author{R.A. Battye}

\address{Department of Applied Mathematics and Theoretical Physics, University of Cambridge, \\ Silver Street, Cambridge CB3 9EW, U.K.}

\maketitle\abstracts{In the light of recent work which suggests that
structure formation by cosmic strings in a critical matter density
universe require large biases, we discuss the effects of reducing the matter density. Using a full linear Einstein-Boltzmann solver, we compute the cold dark matter density field and the microwave background anisotropies/polarization in a flat universe with a non-zero cosmological constant. We find that universes with $\Omega_{\Lambda}=0.7-0.8$ give a good fit to the shape of the observed galaxy distribution, albeit with a bias of around 2. This has interesting implications for the anisotropy and polarization of the microwave background. Finally, we discuss the effects of relaxing the assumption that the universe is flat.}

The last year has seen enormous progress in pinning down the
predictions for structure formation and the microwave background from
defect based theories \cite{PSelTa,ACKDSS,ABRCrit}. When the matter
density is critical, it was found that there is an almost generic
under production of the power on the largest scales currently
observed, with only the most convoluted models achieving a
satisfactory amount of power on $100h^{-1}$Mpc scales, usually at the
expense of an over production on smaller scales. In ref.[3]
this was quantified in terms of the bias required to reconcile the
observed galaxy correlations in the baryonic matter with the
calculated cold dark matter (CDM) power spectrum on those scales. This
was named the $b_{100}$ problem and it was suggested that this bias
could be as large as 5. Such  a bias is both difficult achieve
theoretically and apparently incompatiable with the current
observations in flat, critical density universe.

During the past few months many have described
defect based scenarios as being `ruled out'! While this is very much
human nature and may even be true, it is important to assess the true
scientific implications objectively. A more conservative approach
would be to acknowledge the difficulties these models have in fitting
the current data with the following provisors : (1) {\it Deviations
from scaling are slight.} It is scaling which relates the
contributions on different scales and hence allows for the possibility
of modifying the power spectrum. In ref.[3] an exhaustive
study was made of deviations from scaling and it was concluded that
only truely radical deviations could hope to rectify the the $b_{100}$
problem. (2) {\it Deviations from gaussianity are mild.} Defect based
theories are inherently non-gaussian, but the superposition of a large
number of sources is likely to be gaussian. Since the data is analyzed
under the assumption of gaussianity it may be that we may be being
mislead. However, preliminary indications are that the string models are almost gaussian on $100h^{-1}$Mpc scales \cite{ASWAb}. (3) {\it The data on $100h^{-1}$Mpc is accurate.} On these scales the observed data is dominated by a single survey (APM) and one might suggest that the crucial observations are close to its resolution. Hopefully, future surveys (PSCz, 2DF and SDSS) will improve our confidence in these observations. (4) {\it Large biases are unacceptable.} I have already commented that there is no evidence for large biases, but some are not totally uncomfortable with this possibility. (5) {\it The matter density is critical.} All the work was based on a flat cosmology with standard cosmological parameters, although it was shown that variations in the Hubble constant and baryon density could not solve the $b_{100}$ problem. Since a substantial body of current observations suggest that the universe may have a non-zero cosmological constant or be open, it is important that this possibility is investigated. This is the subject of this review based mainly on ref.[5], but backed up by refs.[6,7,8].  

We will concentrate on the case of a flat universe with a non-zero cosmological constant, since calculations in an open universe using an accurate linear Einstein-Boltzmann solver would require a substantial technological improvement.
However, we will make some qualitative remarks on the open case
later. The reduction of matter density will be seen to have two
related effects on the CDM power spectrum. Firstly, the shift in the
time of equal matter-radiation will effect the shape of the  power
spectrum in the standard way via the shape parameter
$\Gamma\sim\Omega_{\rm m}h$, where $\Omega_{\rm m}$ is the fractional
density in matter and the Hubble constant is $H_0=100h\,{\rm km}\,{\rm
sec}^{-1}\,{\rm Mpc}^{-1}$. In the case of the standard adiabatic
theories, where the perturbations are created around the horizon size,
this is estimated to be $\Gamma\approx 0.2-0.3$. However, the
perturbations are created inside the horizon in defect theories and so
the size of perturbations created at the time of equal
matter-radiation will be slightly smaller, with the consequence that the required value of $\Gamma$ is slightly smaller. We find that $\Omega_{\rm
m}\approx 0.2-0.3$ with $h=0.5$, that is $\Omega_{\Lambda}=0.7-0.8$ and $\Gamma\approx 0.1-0.15$, gives a particularly good fit to the shape of the data for a pure scaling model.

\begin{figure}[t]
\centerline{\psfig{file=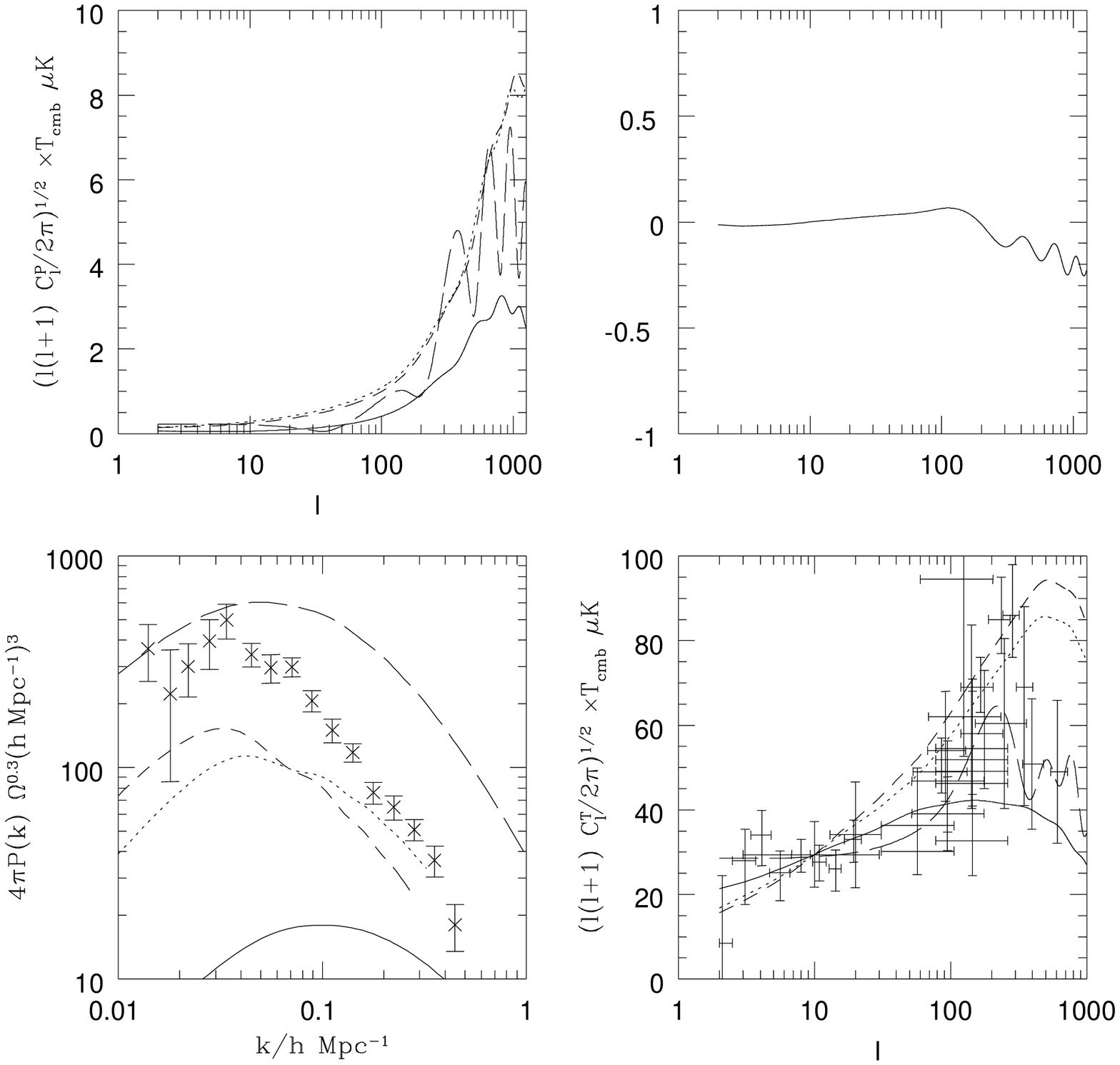,width=2.75in}}
\caption{The results of the calculations for the standard scaling
source (solid line), $\Omega_{\lambda}=0.7$ (dotted line) and
$\Omega_{\lambda}=0.8$ (short-dashed line), all compared with standard
CDM (long-dashed line): (bottom,left) the CDM density field,
(bottom,right) the microwave backgorund anisotropies, (top,left) the
polarization of the microwave background, (top,right) the
cross-correlation of the anisotropy and polarization, relative to the
amplitude of the two components. } 
\end{figure}

The other effect is related to the modifications in the scaling regime
during the matter-radiation transition. It is well known that the
density of strings will be reduce by around a factor of two during
this transition and this can have a substantial influence on the power
spectrum. It has already been pointed out  that radical deviations
from scaling can lead to an improvement in $b_{100}$ for a critical
matter density universe, but smaller deviations can only effect
smaller scales. In a model with a cosmological constant, the time of
equal matter-radiation is shifted in such a way that such a small
deviation can  effect $b_{100}$. Using the velocity dependent
one-scale model \cite{MSa} to model these deviations, we have computed
the CDM power spectrum and the angular power spectra of the micowave
background anisotropies and polarization using a modified version of
CMBFAST \cite{cmbfast}, which the includes the scalar, vector and tensor perurbations generic in any defect model. The results are presented in the figure for $\Omega_{\lambda}=0.7,0.8$ and $h=0.5, \Omega_{\rm b}=0.05$, along with those for pure scaling \cite{ABRCrit} and standard CDM.

The first thing to notice is the substantial improvement in the CDM
power spectrum relative to the data, although  one still requires a
bias of around 2. This is still quite large, but the baryon density is
now substantial proportion of the total and it may be easier to create
a substantial bias, since the baryons are not forced to necessarily
trace the CDM. The angular power spectrum of the microwave background
anisotropies is also effected in a positive way relative to the
current observations. The critical matter density case is not
favourable when compared to the data, with the absence of a
substantive  rise in the power spectrum as suggested, but not yet
confirmed, by the data. However, the models with a non-zero
cosmological constant do have a substantial rise, albeit around
$l=400-600$, rather than $l=200-300$ as suggested by the data. The
reason for this increase in power on smaller scales is that one has
changed the relative normalization between the anisotropy created at
the time of last scattering on small scales and that created along the
line of sight by the integrated Sachs-Wolfe effect (ISW), which
dominate the larger scales, plus the decrease in $\Omega_m$ that is
known to increase the amount of power on smaller scales. It is not a series of peaks as in the adiabatic case, but a broad peak as has been suggested as the signature of an incoherent source \cite{ACFM}. As one final point on this subject, one should note that too much power on small scales is not likely to be a problem when comparing to the data, since it can be removed easily, in this case by reionization. This would also slightly reduce the bias required to fit the galaxy correlations.

The angular power spectrum of polarization and its cross correlation
with the temperature anisotropies are also of interest since future
satellite missions, as well as ground based measurements, may measure
them. There are a number of interesting signatures for active models
which can be seen clearly in this model. Firstly, we should note that
the amplitude of the polarization in these models has a maximum rms of
$8\mu K$ around $l=800-1000$, but that there are two distinct parts to
this, the so called electric, $E$, and magnetic, $B$, components. The
$B$-component is only present in the models where there are tensor and
vector components to the anisotropies and is still relatively small
$(\sim 0.1\mu$K around $l=100$) in adiabatic models with a
substantial tensor component. In the models under consideration here it is much larger, with a maximum rms of around $2\mu$K (note that the standard scaling model \cite{PSelTb} has an maximum rms of $1\mu$K), which may be detectable.

The cross correlation between temperature and polarization on scales
below the scale of the horizon at last scattering is zero by
causality, in contrast to the adiabatic case \cite{ZS}. In particular
it is zero around $l=100$, when the adiabatic model is strongly
anti-correlated. Most interesting, though, from the point of
observations is the suppression of the cross correlation on smaller
scales, which is a characteristic of the incoherent nature of the
source \cite{BH,PSelTb}. At a very basic level the anisotropy and
polarization are two oscillatory functions which are out of phase,
$\cos(x)$ and $\sin(x)$ say. To compute the angular power spectra for
the anisotropy and polarization one must integrate the square of this
oscillatory function multiplied by a structure function, encoding the
precise details of the source. Since this is always positive, one gets
the characteristic broad peak in the spectrum. But when one computes
the cross correlation, it is the product of the two which multiplies
the stucture function and integrates out to be almost
zero. However, one can see the that the cross correlation, although
suppressed, appears to be consistently negative on small scales, that
is, the temperature and polarization are anti-correlated The slighly
more subtle reason for this is that there are actually two oscillatory
components to the anisotropy, from the intrinsic and Doppler
effects, which are responsible for the peaks and troughs in the
standard adiabatic spectrum. The Doppler component is out of phase
and suppressed relative to the intrinsic, but is of the opposite sign and in phase with the source of the polarization, hence creating a negative cross-correlation.

We have shown that some of the problems of standard defect scenarios
in underproducing power in the large scale matter distribution can be
remedied by the reduction of the matter density from
critical. Although we have concentrated on the case of a flat universe
with a non-zero cosmological constant, the general picture that we
have presented will also be true for an open universe \cite{ACM,ASWAa},
with a couple of simple modifications. Firstly, the bias required to
fit the galaxy correlations will be slightly larger. Using the simple
formula of ref.[8], one can estimate that this would increase
by around a factor of 1.4 for $\Omega_{\rm m}=0.2$. Also the CMB power
spectra will be shifted  by the familiar geometric factor, $\Omega_{\rm
m}^{-1/2}$, for adiabatic models. Hence, we conclude that the cosmological constant based models have a better chance of fitting the observations, if the current trends are confirmed. 

\section*{Acknowledgements}
 
I would like to thank Andy Albrecht and James Robinson for their collaboration in the work presented here. I have benefitted substantially from conversations with Neil Turok and particularly Paul Shellard, who first suggested the implications of scaling deviations to me. The computations presented here were done at the UK National Cosmology Supercomputing Centre, supported by PPARC, HEFCE and Silicon Graphics/Cray Research. I am currently supported by Trinity College.

\section*{References}

\def\jnl#1#2#3#4#5#6{\hang{#1, {\it #4\/} {\bf #5}, #6 (#2).}}
\def\jnltwo#1#2#3#4#5#6#7#8{\hang{#1, {\it #4\/} {\bf #5}, #6; {\it
ibid} {\bf #7} #8 (#2).}} 
\def\prep#1#2#3#4{\hang{#1, #4.}} 
\def\proc#1#2#3#4#5#6{{#1 [#2], in {\it #4\/}, #5, eds.\ (#6).}}
\def\book#1#2#3#4{\hang{#1, {\it #3\/} (#4, #2).}}
\def\jnlerr#1#2#3#4#5#6#7#8{\hang{#1 [#2], {\it #4\/} {\bf #5}, #6.
{Erratum:} {\it #4\/} {\bf #7}, #8.}}
\def\prl{Phys.\ Rev.\ Lett.}
\def\pr{Phys.\ Rev.}
\def\pl{Phys.\ Lett.}
\def\np{Nucl.\ Phys.}
\def\prp{Phys.\ Rep.}
\def\rmp{Rev.\ Mod.\ Phys.}
\def\cmp{Comm.\ Math.\ Phys.}
\def\mpl{Mod.\ Phys.\ Lett.}
\def\apj{Ap.\ J.}
\def\apjl{Ap.\ J.\ Lett.}
\def\aap{Astron.\ Ap.}
\def\cqg{Class.\ Quant.\ Grav.} 
\def\grg{Gen.\ Rel.\ Grav.}
\def\mn{MNRAS}
\def\ptp{Prog.\ Theor.\ Phys.}
\def\jetp{Sov.\ Phys.\ JETP}
\def\jetpl{JETP Lett.}
\def\jmp{J.\ Math.\ Phys.}
\def\zpc{Z.\ Phys.\ C}
\def\cupress{Cambridge University Press}
\def\pup{Princeton University Press}
\def\wss{World Scientific, Singapore}
\def\oup{Oxford University Press}

\pagebreak
\pagestyle{empty}

\end{document}